\def\doublespace{\baselineskip=\normalbaselineskip 
\multiply\baselineskip by 2}

\doublespace

\def\frac#1/#2{\leavevmode\kern.1em
\raise.5ex\hbox{\the\scriptfont0 #1}\kern=.1em
/\kern-.15em\lower.25ex\hbox{\the\scriptfont0 #2}}

\documentstyle[preprint,aps]{revtex}

\begin{document}
\draft
\preprint{\today}
\title
{The Imaginary Part of the Optical Conductivity of $\BKBO$}
\def \bk{{\bf k}}
\def \bQ  {{\bf Q}}
\def \bp {{\bf p}}
\def \bq {{\bf q}}
\def\wtilde{\tilde {\omega }}
\def\csqrt{\sqrt{\wtilde^2(\omega) - \phi^2(\omega)}}
\def\YBCO{YBaCu_3O_{7-x}}
\def\BKBO{Ba_{1-x}K_xBiO_3}
\def\bea {\begin{eqnarray}}
\def\eea {\end{eqnarray}}
\def\be {\begin{equation}}
\def\ee {\end{equation}}

\def\bl {{\bf l}}
\def \Dl {\Delta_l}
\author{F. Marsiglio$^{1,2,3}$, J.P. Carbotte$^{2,3}$, A. Puchkov$^2$,
and T. Timusk$^{2,3}$}
\address
{$^1$Neutron \& Condensed Matter Science\\
AECL, Chalk River Laboratories, Chalk River, Ontario, Canada K0J 1J0\\
$^2$Dept. of Physics \& Astronomy, McMaster University, 
Hamilton, Ontario L8S 4M1 \\
$^3 $ Canadian Institute for Advanced Research, McMaster University,
Hamilton, ON L8S 4M1}

\maketitle
\begin{abstract}

The frequency dependence of the imaginary part of the infrared 
conductivity is calculated for a superconductor.
Sharp structure, characteristic of superconductivity with an order parameter
with s-wave symmetry, appears in the BCS limit as a
minimum at a frequency equal to twice the gap value.
This structure scales with
temperature but gets progressively smeared and shifted as impurity
scattering is increased.  The relationship between low
frequency results and the zero frequency limit is investigated.
Experimental results on $\BKBO$ are also presented. The low frequency
imaginary part of the conductivity displays a minimum at $2\Delta \approx
12$ meV, and provides unequivocal evidence of an s-wave superconducting
order parameter.  Strong coupling (Eliashberg) results show similar trends.
Using this formulation we find that the electron-phonon coupling in $\BKBO$
must necessarily be small, with coupling constant $\lambda \approx 0.2$,
in agreement with conclusions drawn from measurements of the
real part of the conductivity. Thus $\BKBO$ is an s-wave superconductor
that is not driven by the electron-phonon interaction.
\end{abstract}
\vfil\eject
\bigskip
\noindent {\bf I.  INTRODUCTION}
\par
Optical conductivity data in the infrared can give important
information on the properties of the low energy excitations present, and,
in particular, on the scattering rate \cite{1}.  In Drude theory, the width
of the real part of the conductivity gives the inverse of the
scattering time, which can be a combination of elastic impurity scattering and
inelastic (e.g. phonon-mediated) scattering.  In
the latter case, a new process, phonon-assisted
absorption, becomes possible \cite{2}.  This process does not exist with
impurity scattering alone, and makes an additional contribution to
both the real and imaginary parts of the conductivity.

     In the past most discussions of the infrared conductivity in
an electron-phonon system have concentrated on the real part
of the conductivity \cite{1}.  Very few results are available on the
imaginary part which nonetheless follows uniquely from the real
part through a Kramers-Kronig relation.  Recently
experimentalists have effectively exploited the relationship
between the inverse square of the penetration depth and the zero
frequency limit of the imaginary part of the conductivity 
($lim_{\nu \rightarrow 0} \nu \sigma_2(\nu) = {c^2 \over 4\pi}{1 \over
\lambda^2(T)}$, where we use the convention that $\sigma = \sigma_1 - 
i\sigma_2$, and $\lambda$ here is the penetration depth), to obtain information on this important
quantity from the far-infrared optical conductivity \cite{3}.\par

The purpose of this paper is to study the frequency dependence 
of $\nu \sigma_2(\nu)$. We
start with the BCS model \cite{bcs} and then generalize to an Eliashberg 
strong coupling
system.  
The aim is to understand how the more complicated
effects due to the interactions with phonons modify the BCS results.
The necessary theory is
summarized in section II.  Results of BCS theory are presented in
section III.  It is found that in an s-wave superconductor there
will be a distinct minimum in $\nu \sigma_2(\nu)$ at twice the gap value.
With the addition of impurities the minimum broadens and shifts slightly.
The temperature dependence of the minimum tracks the gap temperature
dependence.  We also study how
finite frequency results are related to the zero frequency limit.
Results are presented in section IV for the strong coupling case.
The most significant difference between the Eliashberg and
BCS approaches occurs in the pure case.  In this limit, BCS theory
no longer contains any scattering mechanisms, whereas
inelastic scattering remains in an Eliashberg
formalism, and we can study the effect of this additional source of
scattering.  In section V, the specific case of $\BKBO$ (with $x = 0.4$) will be
considered. Comparison with infrared data will lead us
to conclude that this system contains an isotropic s-wave order parameter,
but is {\it not} an electron-phonon superconductor.
The inelastic scattering observed
in the conductivity at room temperature is inconsistent with
a mass enhancement parameter of order unity,  which is required
for an electron-phonon superconductor with $T_c \approx 30$ K.
A summary is provided in section VI.
\par
\bigskip
\noindent {\bf II.  FORMALISM}
\par
Expressions for the optical conductivity of an electron-phonon
superconductor, excluding 
vertex corrections for the electron-phonon interactions, are now
well established \cite{4,5,6,7,8,9,10,11,12} and 
require solutions of the Eliashberg
equations \cite{13} for the pairing function $\phi(\omega)$ and renormalized 
frequency $\wtilde(\omega)$.
For an isotropic s-wave
electron-phonon superconductor, these equations take the form \cite{14}
\begin{eqnarray}
\phi(\omega) & = & \pi T\sum\limits_{m=-\infty}^{\infty}
\bigl[\lambda(\omega-i\omega_m)-\mu^*(\omega_c)\theta(\omega_c-\vert{\omega_m}
\vert)
\bigr]{\phi_m\over\sqrt{\wtilde^2(i\omega_m)+\phi_m^2}}
\nonumber \\
 & & +i\pi\int_0^{\infty}d\nu\,\alpha^2F(\nu)\Biggl\{\bigl[N(\nu)+f(\nu-
\omega)\bigr]{\phi(\omega-\nu)\over\sqrt{\wtilde^2(\omega-\nu) -
\phi^2(\omega-\nu)}}
\nonumber \\
 & & +\bigl[N(\nu)+f(\nu+\omega)\bigr]{\phi(\omega+\nu)\over\sqrt{
\wtilde^2(\omega+\nu) - \phi^2(\omega+\nu)}}\Biggr\}
\label{phireal}
\end{eqnarray}
\noindent and
\begin{eqnarray}
\wtilde(\omega) & = & 1 + i\pi T\sum\limits_{m=-\infty}^{\infty}
\lambda(\omega-i\omega_m){\wtilde(i\omega_m)\over\sqrt{\wtilde^2(i\omega_m)
+\phi_m^2}}
\nonumber \\
 & & + i\pi \int_0^{\infty}d\nu\,\alpha^2F(\nu)\Biggl\{\bigl[N
(\nu)+f(\nu-\omega)\bigr]{\wtilde(\omega-\nu)\over\sqrt{\wtilde^2(\omega-\nu)
-\phi^2(\omega-\nu)}}
\nonumber \\
& & +\bigl[N(\nu)+f(\nu+\omega)\bigr]{\wtilde(\omega+\nu)\over\sqrt{
\wtilde^2(\omega+\nu)-\phi^2(\omega+\nu)}}\Biggr\}\ .
\label{wtildereal}
\end{eqnarray}
In these equations $N(\nu)$ and $f(\nu)$ are the Bose
and Fermi distribution functions, respectively, and
$\lambda(z)$ is the Hilbert transform of the electron-phonon spectral
function $\alpha^2F(\nu)$. $\mu^\ast(\omega_c)$ is the Coulomb repulsion
parameter with cutoff $\omega_c$. A negative value for this parameter
can be used to model some BCS attraction of unspecified origin. The required
functions are first obtained on the imaginary axis at the 
Matsubara frequencies, i.e. $\omega = i\omega_n \equiv
i\pi T(2n-1)$, with $\phi_m \equiv \phi(i\omega_m)$ by setting the complex
variable $\omega$ in these equations to the Matsubara frequencies \cite{15,16}.
Then the equations are iterated as written, with $\omega$ set to a frequency
on the real axis.\par

   From the numerical solutions of Eqs. (\ref{phireal}-\ref{wtildereal}), the
conductivity follows from the paramagnetic response kernel,
$\Pi(\nu + i\delta)$, given by \cite{4,5,7,8,9,10,11,12}
\begin{eqnarray}
\Pi(\nu + i\delta) & = & {ne^2 \over m}
\Biggl\{ -1 + \int_0^\infty d\omega \tanh({\beta\omega \over 2})
\Bigl(
h_1(\omega,\omega + \nu) -  h_2(\omega,\omega + \nu)
\Bigr)
\nonumber \\
& & \phantom{{ne^2 \over m}\Biggl\{ -1} + \int_{-\nu}^D
d\omega \tanh({\beta (\omega + \nu) \over 2})
\Bigl(
h_1^\ast (\omega,\omega + \nu) +  h_2(\omega,\omega + \nu)
\Bigr)
\Biggr\}
\label{kernel}
\end{eqnarray}
\noindent with the conductivity
\begin{equation}
\sigma(\nu) = {i \over \nu} \Bigl( \Pi(\nu + i\delta) + {ne^2 \over m} \Bigr),
\label{cond}
\end{equation}
\noindent where e is the magnitude of the charge on the electron, m its mass
and n the number of electrons per unit volume.  The functions $h_i$ in
Eq. (\ref{kernel})
are given by
\begin{eqnarray}
h_1(\omega_1,\omega_2) & = & {1 - N(\omega_1) N(\omega_2) -
P(\omega_1) P(\omega_2) \over 2(\epsilon(\omega_1) + \epsilon(\omega_2)) }
\nonumber \\
h_2(\omega_1,\omega_2) & = & {1 + N^\ast(\omega_1) N(\omega_2) +
P^\ast(\omega_1) P(\omega_2) \over 2(\epsilon(\omega_2) -
\epsilon^\ast(\omega_1)) }
\nonumber \\
N(\omega) & = & { \wtilde(\omega +i\delta) \over \epsilon(\omega + i\delta) }
\nonumber \\
P(\omega) & = & { \phi(\omega +i\delta) \over \epsilon(\omega + i\delta) }
\nonumber \\
\epsilon(\omega) & = &
\sqrt{\wtilde^2(\omega +i\delta) -\phi^2(\omega +i\delta )}.
\label{definitions}
\end{eqnarray}
\noindent The equations so far have been written for a clean system.  To
include normal impurity scattering in the Born approximation, an
average scattering rate, $1/\tau$, is used, and gives rise to an
additional contribution to the pairing and renormalization functions:
\begin{equation}
\phi(\omega) \rightarrow \phi(\omega) + {i \over 2\tau} {\phi(\omega)
\over \csqrt}
\label{phidirt}
\end{equation}
\begin{equation}
\wtilde(\omega) \rightarrow \wtilde(\omega) + {i \over 2\tau} {\wtilde(\omega)
\over \csqrt}.
\label{wtildedirt}
\end{equation}
\noindent Eq. (\ref{kernel}) remains the same with impurity scattering.
The modifications are all implicitly contained in the pairing and
renormalization functions. Note that the gap parameter, $\Delta(\omega)
\equiv \phi(\omega)\bigg/ Z(\omega)$, remains the same, independent of 
the impurity scattering rate.
\par
\bigskip
\noindent {\bf III.  THE BCS LIMIT}
\par
    The equations for the conductivity and pairing function discussed
so far reduce in the BCS limit, thus simplifying the numerical work 
\cite{17,18,19}.
The result is a universal function for the conductivity, dependent only
on the impurity scattering rate. In the BCS limit $\wtilde(\omega)$
is given by the bare frequency $\omega$ (plus the impurity scattering
term given in Eq. (\ref{wtildedirt})), and the pairing function
$\phi(\omega)$ reduces to the BCS gap parameter, $\Delta$ (plus
Eq. (\ref{phidirt})). The temperature
dependence of $\Delta(T)$ is given by solution of the BCS equation. As already
mentioned,
the clean limit is pathological within BCS theory. To allow
some scattering mechanism to remain some nonzero impurity scattering
is required. Note that in the remainder of this paper we present
results
for the conductivity normalized to $ne^2/m$, so the only parameter
in the BCS limit is the impurity scattering rate, $1/\tau$.
\par
     Results are shown in Fig. 1  for $\nu \sigma_2(\nu)$
as a function of frequency for a BCS model with $T_c = 29$ K
($\BKBO$). We use  (a) $1/\tau = 1$, (b) $1/\tau = 5$, and 
(c) $1/\tau = 25$ meV, with differently styled curves for various
temperatures, as explained in the figure caption.
The largest scattering rate we have used is representative
of the $\BKBO$ compounds, as we will see later.

For small scattering rate (Fig. (1a)) the quantity $\nu \sigma_2
(\nu)$ displays a cusp-like minimum at a frequency equal to two
times the temperature dependent gap parameter. As the temperature
increases towards $T_c$ the size of the minimum decreases, and
of course its location tracks the energy gap. For $1/\tau = 1$ meV
the cusp-like minimum occurs at perhaps too low a frequency for
temperatures near $T_c$ to be observed with optical infrared
techniques, but at least in principle this minimum can be used
to measure the gap temperature dependence. Note that the cusp in
the imaginary part of the conductivity simply reflects the
sharp onset in the real part of the conductivity which occurs at
$2\Delta(T)$. For larger impurity scattering rates the minimum
broadens somewhat.\par

Another important
feature of our results is the limit as the frequency goes to
zero, which as stated earlier, gives the London penetration depth.
We have verified that our low frequency conductivity results
extrapolate to the correct penetration depth, as computed from an
imaginary axis formulation \cite{4,20}:
\begin{equation}
{c^2 \over 4\pi} {1 \over \lambda^2(T)} = {ne^2 \over m}
\pi T \sum_{n = -\infty}^{\infty} {\phi^2(i\omega_n) \over
(\wtilde^2(i\omega_n) + \phi^2(i\omega_n))^{3/2}}.
\label{penet}
\end{equation}
\noindent As is apparent from the figures, particularly for
low scattering rate, this limit can only be probed with very low
photon energy, certainly too low for the usual infrared range in
conventional electron-phonon superconductors.
\par

As the impurity scattering rate increases the low frequency
conductivity $\nu \sigma_2(\nu)$ deviates significantly from unity.
This
is expected since the penetration depth increases with
impurity scattering in a BCS model. The effect is very pronounced in
Fig. (1c) where it is also seen that the minimum due to the gap is
considerably broadened due to the impurity scattering. To see the
effect of impurity scattering on the low frequency conductivity more
clearly we show in Fig. 2 results for various impurity scattering
rates. All curves are calculated in the BCS limit and for reduced
temperature, $T/T_c = 0.1$, which is essentially zero temperature.
As the scattering rate increases, the cusp-like minimum broadens
into a shallow minimum. For very high scattering rates, it is evident
that extracting a gap value would be very difficult. The actual minimum
is displaced to higher frequencies, and is no longer located at
twice the gap.
\par
The results of Figs. (1-2) immediately suggest the following question:
To what extent can finite frequency conductivity data be used to
extract information about the zero frequency penetration depth ? 
We first address the question of temperature dependence.
The dimensionless quantity $\nu \sigma_2(\nu)$ is plotted in Fig. 3
as a function of reduced temperature, for various frequencies, as
given in the figure caption. Two impurity scattering rates are used,
(a) $1/\tau = 1$ meV to represent a fairly clean case, and (b)
$1/\tau = 25 $ meV, to represent both a dirtier case and $\BKBO$.
In Fig. (3a) the lowest frequency curve plotted ($\nu = 0.2$ meV)
does indeed follow very closely the temperature dependence of the
zero frequency penetration depth. Nonetheless, already at $\nu = 1.0
$ meV (dotted curve) the temperature dependence of $\nu \sigma_2(\nu)$
deviates very markedly from that of the solid curve, and hence
cannot be used to extract the temperature dependence of the penetration
depth. As the impurity scattering increases the finite frequency
results improve  as is seen in Fig. (3b). Clearly the finite frequency
results give a good qualitative estimate of the temperature
variation of the penetration depth up to $\nu = 2.5$ meV. 
For frequencies of order twice the gap (e.g. $\nu =  10$ meV),
however, all
correspondence with the penetration depth is lost.
\par

Can finite frequency data be used to extract the
impurity dependence of the penetration depth ?
In the BCS limit, with $\alpha \equiv {1 \over 2\Delta \tau}$
\cite{21},
\bea
{1 \over \lambda^2(T=0)} & = & {1 \over \lambda_{cl}^2(T=0)} \Biggl\{
{\pi \over 2 \alpha} - {1 \over \alpha \sqrt{1 - \alpha^2}}\sin^{-1}
(\sqrt{1 - \alpha^2}) \Biggr\} \phantom{bbbbbbbbbbbbbb} \alpha < 1
\nonumber \\
& & {1 \over \lambda_{cl}^2(T=0)} \Biggl\{
{\pi \over 2 \alpha} - {1 \over 2 \alpha \sqrt{\alpha^2 - 1}}\ln
\Bigl( {\alpha + \sqrt{\alpha^2 - 1} \over a - \sqrt{\alpha^2 - 1} }
\Bigr) \Biggr\} \phantom{bbbbbbbbbb} \alpha > 1.
\label{londirt1}
\eea
\noindent In the weak scattering limit this expression reduces
to the more familiar form,
\be
{1 \over \lambda^2(0)} \approx {1 \over \lambda_{cl}^2(0)} {1 \over
1 + {\pi \over 4} \alpha}.
\label{londirt2}
\ee
\noindent This expression can be written in terms of the
zero temperature coherence length, $\xi_0$, and the mean free path,
$\ell$, using $\Delta = {v_F \over \pi \xi_0}$ and $v_F = \ell/\tau$,
where $v_F$ is the Fermi velocity: 
\be
{1 \over \lambda^2(0)} \approx {1 \over \lambda_{cl}^2(0)} {1 \over
1 + {\pi^2 \over 8} {\xi_0 \over \ell}},
\label{londirt3}
\ee
\noindent In Fig. 4 we plot $\nu \sigma_2(\nu)$ at various
frequencies as a function of the scattering rate, $1/\tau$. Clearly,
up to about 5 meV the deviations from the zero frequency curve,
Eq. (\ref{londirt1}), are small. Significant deviations begin to
occur at higher frequencies ($\nu = 10 $ meV). Thus, one must be
careful when using finite frequency data to infer information
about the zero frequency penetration depth.
\par
\bigskip
\noindent {\bf IV.  THE STRONG COUPLING CASE}
\par

     The results presented so far are based on BCS theory, where
the interaction is modelled by a constant, structureless in momentum
and frequency (except for a cutoff).  One of the merits of such an
approach is that the results are universal, but the disadvantage is that
inelastic scattering is not included and so strong coupling
effects in the superconducting state are not accounted for.  Strong
coupling results first require a full numerical
solution to the Eliashberg equations (\ref{phireal},\ref{wtildereal}).
Two parameters are needed to characterize a particular material,
the electron-phonon spectral function $\alpha^2F(\omega)$, and
the Coulomb repulsion, $\mu^\ast$. The latter is often fitted to
either $T_c$ or the gap, while the former is extracted through
an inversion procedure from tunnelling data \cite{22}. Often 
$\alpha^2F(\omega)$ has been found
to be given roughly by  a constant times the phonon frequency
distribution, $F(\omega)$. 
Thus, for most purposes, the difference between $\alpha^2F(\omega)$
and $F(\omega)$ 
can be ignored and a single multiplicative parameter can be
used to obtain $\alpha^2F(\omega)$ from $F(\omega)$
when tunnelling data is not
available.  To be specific here, we will treat only the case of
$\BKBO$. Its frequency distribution is known \cite{23} and is similar to
calculated \cite{24} values of $\alpha^2F(\omega)$ as well as 
values determined from
tunnelling \cite{25}. Based on these three spectra then if 
$\BKBO$ is an electron-phonon superconductor,
the value of the electron mass renormalization
$\lambda = 2\int d\omega \alpha^2F(\omega)/\omega$ needed to produce 
superconductivity with $T_c \approx 30$ K is $\lambda \approx 1$. 
Because of difficulties with the inversion of the tunnelling data \cite{25},
we will use a spectrum scaled from neutron scattering results
with $\lambda = 1$ as a representative intermediate coupling
Eliashberg superconductor. In fact Sharifi {\it et al.} have argued, based
on tunnelling measurements, that $\BKBO$ is not an electron-phonon
based superconductor \cite{sharifi}. In the next section we will 
also argue that 
$\BKBO$ cannot be an electron-phonon superconductor and that
$\lambda$ is
more likely to be of order 0.2, a value much smaller than considered in the
work previously quoted \cite{23,24,25}. \par

     In Fig. (5a) we show results for the product of
the frequency times the imaginary part of the conductivity, i.e.
$\nu \sigma_2(\nu)$,  as a function of $\nu$ for a strong 
coupling superconductor with
$\lambda = 1$ and $T_c= 29$ K  as discussed above.  The curves are plotted in
the clean limit, i.e. $1/\tau = 0$. Scattering is still possible
through inelastic scattering via the phonons, which are present at finite
temperature.
At low temperatures a very sharp Drude-like
peak will be present in the normal state conductivity.  This manifests
itself in the imaginary part of the conductivity as a sharp inverted
Drude-like minimum near the origin \cite{26}, as is visible in the lower
left portion of Fig. (5a).
The solid curve is for $T/T_c = 0.99$ while the dotted curve,
which is almost indistinguishable from the solid curve,
is for the normal state at the same temperature. This curve extends down to
zero at the origin (not shown) and has a width of order 1 meV. Also shown
are results at $T/T_c = 0.5$ (dashed curve) and $T/T_c = 0.2$ (dot-dashed
curve). These curves are very different from their counterparts in the previous
section. The first thing to note is that sharp structure at $\nu = 2\Delta
\approx 10.4 $ meV is completely absent. In this case there is no BCS reference,
as in this limit the corresponding result would be a constant at value
unity (see Fig. 1). Recall that in the real part of the conductivity absorption
begins beyond $2 \Delta$ not in an abrupt fashion as in the Mattis-Bardeen
limit of the theory, but smoothly, roughly following the phonon density
of states since the phonons conserve the energy and account for the momentum
of the created electron-hole pair \cite{7,27}.
Since the absorption edge
is not abrupt in $\sigma_1(\nu)$, we cannot expect the corresponding sharp
structure in the imaginary part $\nu \sigma_2(\nu)$; this is born out by our
calculations shown in Fig. (5a). \par

Another difference which is obvious upon inspection of Fig. (5a) is that
structure is present in the imaginary part of the conductivity over a
frequency range representative of the phonon energies in the 
$\alpha^2F(\omega)$ spectrum. They occur here in much the same way
as in the real part of the conductivity \cite{4,12}, and
leave open the possibility for the determination of $\alpha^2F(\omega)$
by infrared spectroscopy. \par

Finally another clear difference occurs in the zero frequency limit,
where, at low temperatures, the intercept is approximately 0.5, i.e.
half of the London limit. This is due to phonon renormalization
effects, which, for the most part amount to a $1 + \lambda$ enhancement
of the mass, as is obvious from Eq. (\ref{penet}) \cite{28}.\par

For non-zero impurity scattering two important things happen, as Fig. (5b)
shows. A cusp-like minimum, conspicuous by its absence in the clean limit,
is now present even with a small impurity scattering rate, $1/\tau = 2$ meV.
Secondly the penetration depth, given by the inverse of the zero frequency
limit of the quantity plotted (see remark in the introduction), has increased
with the addition of impurities, as discussed in the previous section
in the BCS limit. The increase here is non-universal, i.e. it depends on the
strength of the interaction, and somewhat less sensitively on
the form of the underlying electron-phonon spectral function. \par

To see the effect of coupling strength on the low frequency imaginary
part of the conductivity, the quantity $\nu \sigma_2(\nu)$ is plotted
in Fig. 6 vs. frequency with (a) $1/\tau = 2 $,
and (b) $1/\tau = 25 $ meV. The curves plotted are for various coupling
strengths as indicated, with the Coulomb repulsion $\mu^\ast$ adjusted
in each case so that $T_c$ remains 29 K. We have used a small
Coulomb repulsion to model $\BKBO$ with $\lambda = 1$ so that for
small $\lambda$ we use negative $\mu^\ast$'s, thus representing some
additional attractive mechanism by which the electrons bind.
In Fig. (6a) there is a clear lowering of the imaginary part of the
conductivity with increasing coupling strength, while in Fig. (6b)
the value of $\nu \sigma_2(\nu)$ is drastically reduced even in the
BCS limit ($\lambda = 0$). In both cases the depth and sharpness of the
minimum associated with $2\Delta$ decreases as the coupling strength increases.
The minimum also shifts slightly to higher frequency, as expected since the
zero temperature gap increases with increasing coupling strength. At higher
frequencies phonon structure becomes more noticeable as coupling strength
increases.\par

     In Fig. 7 we show similar results to those given in Fig. 5
but now with an impurity scattering rate equal to
50 meV.  Upon comparison with Fig. 5, it is clear that the entire
curve for $\nu \sigma_2(\nu)$ at low temperatures ($T/T_c = 0.2$)
has been reduced
systematically by the impurity scattering. Furthermore a gap minimum
is clearly present, as Fig. (5b) would suggest.
Phonon structure is still seen at higher energies.  Also, the value
of the penetration depth has been reduced significantly in a
fashion that roughly obeys Eq. (\ref{londirt1}).  There are strong
coupling corrections \cite{20} so that exact agreement is not
expected. \par

 The higher temperature curves are also quite interesting
and deserve some attention.  The solid curve is for the
superconducting case with $T/T_c = 0.99$ and falls almost exactly on the
dotted curve which is the normal state result at the same temperature
(differences are only apparent at very low frequency).
In the normal state, the low frequency part of the real and
imaginary part of the conductivity can be fit to a Drude-like contribution, as
first attempted by Dolgov et al. \cite{29} and taken up more
microscopically and rigorously recently by us \cite{12}.  If both real and
imaginary parts are fitted in a low frequency expansion of the
normal state conductivity, we arrive at the result
\begin{equation}
\sigma_{Drude} \equiv {ne^2 \over m}{1 \over 1 + \tilde{\lambda} }
{  \tilde{\tau} \over 1 - i\nu \tilde{\tau}  },
\label{drude}
\end{equation}
\noindent with
\def\tt{\tilde{\tau}}
\def\lt{\tilde{\lambda}}
\begin{equation}
\tt \equiv { \Biggl\langle \bigl(\tau^\ast(\omega)\bigr)^2/\Bigl(1 + 
\lambda^\ast(\omega)\Bigr) \Biggr\rangle \over 
\Biggl\langle \tau^\ast(\omega)/\Bigl(1 + 
\lambda^\ast(\omega)\Bigr) \Biggr\rangle },
\label{tautilde}
\end{equation}
and
\begin{equation}
1 + \lt \equiv \tt \Bigg/ \Biggl\langle { \tau^\ast(\omega) \over 1 +
\lambda^\ast(\omega) } \Biggr\rangle ,
\label{lambdatilde}
\end{equation}
with
\begin{equation}
1 + \lambda^\ast(\omega) \equiv {\partial \hbox{Re} \wtilde(\omega) \over
\partial \omega}
\label{lambdaomega}
\end{equation}
and
\begin{equation}
1/\tau^\ast(\omega) \equiv {1/\tau + 2 \hbox{Im} \wtilde(\omega) \over
1 + \lambda^\ast(\omega)}.
\label{tauomega}
\end{equation}
\noindent Note that $\bigl\langle Q(\omega) \bigr\rangle \equiv
\int_0^\infty {\beta d\omega \over 2}
\hbox{sech }^2 \bigl( {\beta \omega \over  2 } \bigr) Q(\omega) $ and
explicit closed expressions 
for both these last quantities are
given in our previous paper \cite{12}. \par

     In Fig. 8 we compare our Drude fits (dotted curves) 
with the full electron-phonon results (solid curves) in the clean
limit as a function of frequency for various temperatures in the
normal state.
At low temperature, i.e. $T/T_c = 1$, 
the Drude fit at low  frequency is very narrow,
reflecting the fact that the electron-phonon scattering time
becomes very long at low temperature.  As the frequency increases
the two curves
begin to deviate very markedly.  In units of $ne^2/m$, $\nu
\sigma_2(\nu)$ as modelled by 
the Drude form Eq. (\ref{drude}) will saturate at high frequencies
to a value of
$1/(1+\lt)$,  which is 0.43 for the chosen parameters.  This saturation is
seen very clearly in the plots.  As the temperature is increased to
$ T/T_c = 5$, the inelastic scattering increases significantly. The solid
curve reflects this fact in that the low frequency variation is on a much
larger scale,  of order 60 meV.  Also, the Drude-like curve (dotted) fits 
the full numerical solution (solid curve)
over a larger frequency range and saturates at a larger value of
$1/(1+\lt)$,  which has increased considerably with increasing temperature
as described in our previous work \cite{12}.  
Note that $\lim_{\nu \rightarrow \infty} \nu \sigma_2(\nu) = 1$ in 
the full calculation (in units of $ne^2/m$).
Thus, one could use $\nu \sigma_2(\nu)$ data to determine  $\lt$ and
$\tt$
independently from the low frequency fit and from the infinite frequency
limit. Finally, the last set of curves is for $T/T_c = 10$. The fit
is very good at low frequency. The two curves cross just beyond
the highest frequency shown, and large deviations occur there. \par

     Returning to Fig. 7, the short-dashed-long-dashed
curve is given for $T/T_c=10$ (290 K).
The width of the imaginary part of the
conductivity has increased drastically over its value for $T/T_c = 1$.
This increase is due entirely to the increased inelastic scattering at
elevated temperatures.  The
total width, of course, is a combination of the inelastic and elastic
contribution; this is clear from a comparison of the $T/T_c=1$ curves of
Figs. 5 and 7.  In Fig. 5, the width of the imaginary part of the
conductivity is very narrow because it is only due to inelastic
scattering while in Fig. 7 it is almost entirely due to the elastic
scattering provided by a $1/\tau = 50$ meV impurity term.  While one might
think that a lower bound for this width is 50 meV, one must recall that
the effective elastic scattering rate is renormalized by the electron-phonon
interaction. It is well known \cite{28,30} that 
the electron-phonon interaction renormalizes $1/\tau$ 
by a factor $1/(1 + \lambda)$, which is about 0.5 in our
calculations. This is why the effective width in Fig. 7 appears to be smaller
than 50 meV, and close to 25 meV.
\par
\bigskip
\noindent {\bf V.  THE SPECIFIC CASE OF $\BKBO$}
\par

     Puchkov et al. \cite{31,32} have obtained experimental results for the
conductivity in samples of $\BKBO$ with varying $x$
in the superconducting region of the phase
diagram.  Based on an analysis of the real part of the conductivity,
the authors of Ref.\cite{31}
conclude, contrary to the conclusions of many other workers \cite{23,24,25},
that this material cannot be a conventional electron-phonon superconductor
with coupling constant value $\lambda \approx 1$. This 
conclusion was reinforced by
the more precise microscopic analysis provided by two of us \cite{12}.
Here, we extend our theoretical analysis to the
imaginary part.  In Fig. 9 we plot the quantity $\nu \sigma_2(\nu)$
as a function of frequency at two different temperatures, one in the
superconducting state, and one in the normal state. The experimental
spectra were obtained using a Kramers-Kronig analysis of the
reflectivity spectra measured from 25 cm$^{-1}$ to 40000 cm$^-1$
as was reported elsewhere \cite{31}. To improve the accuracy of the
analysis, ellipsometric measurements were performed on the same
crystal at frequencies from 10000 to 50000 cm$^{-1}$, and
the high frequency reflectivity approximation was chosen in such a way
that the calculated high frequency optical conductivity was in
agreement with the direct ellipsometric results.
Note that we have used the data at 1500 cm$^{-1}$ (185 meV) 
to normalize the curves.
\par

  The most striking feature of the data in Fig. 9 is the well-defined
minimum which occurs at approximately 12 meV. In light of the discussion
thus far we view this result as unequivocal evidence of an s-wave order
parameter in the $\BKBO$ system. Similar results for other dopant
concentrations (and hence $T_c$'s) are also obtained \cite{32}.
This conclusion is reinforced by complementary work by Jiang and
Carbotte \cite{33} wherein they investigate  the behaviour of the
imaginary part of the conductivity for a superconductor with order parameter
of d-wave symmetry. They find that no minimum occurs in $\nu \sigma_2(\nu)$
at $2\Delta$, in agreement with experimental results on $\YBCO$, which
is generally believed to be a d-wave superconductor. This is 
what is expected since the real part of the conductivity
does not have an abrupt onset at $2 \Delta$, which we have argued
accompanies the minimum in the imaginary part via a Kramers-Kronig transform.
\par

  Given the normalization of the experimental data discussed above,
Fig. 2 makes it clear that a considerable amount of impurity scattering is
required to achieve a qualitative agreement between theory and
experiment. This observation is also consistent with the analysis of
the real part of the conductivity, although in that instance a
``mid-infrared contribution'' of unknown origin had to be subtracted 
from the data first. Then Fig. (6b) shows that for $\lambda$ of
order unity, the large amount of inelastic scattering present at
relatively high frequencies ($\approx 25-50 $ meV) will reduce the
imaginary part of the conductivity, in disagreement with experiment.
This disagreement will be even more significant at higher temperatures.
\par

   Theoretical curves are also plotted in Fig. 9 to more clearly
illustrate these remarks. As a ``good'' representative fit we
show results at the two temperatures using a reduced $\alpha^2F(\omega)$
for $\BKBO$ with $\lambda = 0.2$ (dashed curves). 
We have used $1/\tau = 25$ meV, a value which is consistent with that
obtained experimentally from $\sigma_1(\nu)$ \cite{31,32}.
It is clear that the fit is quantitatively good. For comparison
we also show the result using the full spectrum, i.e. with $\lambda = 1$.
It is obvious that such a fit is poor, and given our previous results,
the reader can appreciate that
no amount of parameter adjustment will yield good agreement with the
data {\it at both temperatures} while retaining $\lambda $ of order unity.
This general result is, of course, consistent with the conclusion inferred
from the real part of the conductivity \cite{31,12}. The analysis here, however,
has the advantage that no data subtraction is required before the analysis.
It has the disadvantage that a high frequency scale is required
to normalize the experimental data shown in Fig. 9.
\par
\noindent {\bf VI.  CONCLUSIONS}
\par

     We have studied the frequency dependence of the 
imaginary part of the conductivity $\nu \sigma_2(\nu)$ for an s-wave
superconductor. Universal BCS results are first established which show
a sharp cusp-like minimum in $\nu \sigma_2(\nu)$ at
twice the gap value. This provides a very good method for determining the
gap value and its temperature dependence.
As the rate of elastic impurity scattering 
is increased, the gap structure becomes progressively smeared
and shifts to higher frequency. This shift is completely uncorrelated 
with the gap value
since the gap is independent of elastic impurity scattering in
an isotropic s-wave superconductor.  In the presence of impurity
scattering strong coupling effects
modify the results by smearing the gap minimum, and renormalizing the
low frequency behaviour through a renormalization $1/\tt \rightarrow
{1/\tau \over 1 + \lambda}$ \cite{34}. In addition, structure 
is apparent in the phonon
region. These are all quantitative changes. In contrast, qualitative
changes occur in the clean limit, since then strong coupling effects become
the only source for electron scattering.
In this case, there is
no structure in  $\nu \sigma_2(\nu)$ at twice the gap value because absorption
across the gap can only proceed through a phonon-assisted mechanism
needed to absorb the necessary momentum and energy.  This means that the real
part of the conductivity will become non-zero at a frequency
of twice the gap plus the lowest phonon energy in the system.
The phonon spectrum generally extends to zero frequency  but the 
density of states for the phonons usually has an $\omega^2$
dependence at low frequency.
This leads to a rather gradual increase in absorption
and no sharp structure in $\nu \sigma_2(\nu)$.  When impurities are added, a
qualitative change occurs because impurity-assisted absorption can
now set in quite abruptly at $2 \Delta$.  This manifests itself in the
imaginary part of the conductivity as a sharp cusp-like minimum in
$\nu \sigma_2(\nu)$.
\par

     Application of our calculations to the specific case of $\BKBO$
allows us to conclude, from a consideration of data on the imaginary
part of the conductivity alone, that this system is an
s-wave superconductor. On the other hand further analysis
suggests that the superconductivity is not phonon-mediated as others
have concluded.  The data is consistent with an electron-phonon 
mass enhancement parameter of about $\lambda \approx 0.2$.
Larger values of order unity suggested in the
literature lead to much too large a scattering rate at room
temperature.
\par

     In summary we believe that measurement of the imaginary part
of the conductivity is a powerful alternative to determine both the 
symmetry and the value of the superconducting order parameter. Used
with normal state measurements, such a probe also allows us to infer the
strength of the inelastic scattering present in a particular
superconductor.  In contrast, in a d-wave superconductor, there will be no gap
minimum at $2 \Delta$ because in such a case there is an entire
distribution of gaps,  starting at value zero. The conductivity
averages over these gaps, and this leads to smooth behaviour in
$\nu \sigma_2(\nu)$.
\par
\noindent {\bf   Acknowledgements}
\par
     This research was partially supported by the Natural Sciences
and Engineering  Research Council of Canada (NSERC) and by the
Canadian Institute for Advanced Research (CIAR).  We thank Dimitri
Basov for helpful discussions.

\par 

\vfil\eject
\bigskip
\noindent {\bf Figure Captions}
\par
\begin{itemize}

\item Fig. 1.  
\def\s2{\nu \sigma_2 (\nu)}
The quantity $\s2$ (imaginary part of the conductivity times the frequency
in units of $ne^2/m$) vs. frequency for various impurity scattering rates,
(a) $1/\tau = 1$, (b) $1/\tau = 5$, and (c) $1/\tau = 25$ meV.
These results are calculated in the BCS limit with $T_c =29$ K.
The zero temperature gap value is given by ${2\Delta_0 \over k_B T_c} = 
3.53$ ($\Delta_0 = 8.8 $ meV). The various curves correspond to
different temperatures: $T/T_c = 0.1$ (long-dashed),
$T/T_c = 0.5$ (long-dashed-short-dashed),
$T/T_c = 0.7$ (long-dashed-dotted),
$T/T_c = 0.8$ (short-dashed),
$T/T_c = 0.9$ (dotted), and $T/T_c = 0.995$ (solid).

\item Fig. 2.
$\s2$ vs. frequency for various impurity scattering rates.
These results are in the BCS limit with $T/T_c = 0.1$.
The scattering rates are $1/\tau= 1$ meV (solid), 
$1/\tau= 5$ meV (dotted), $1/\tau= 10$ meV (dashed), 
$1/\tau= 25$ meV (dot-long-dashed), $1/\tau= 50$ meV (short-dashed-long-dashed),
and $1/\tau= 100$ meV (long-dashed). 

\item Fig. 3.
$\s2$ vs reduced temperature $T/T_c$ in the BCS limit, plotted for various
frequencies. The results are for (a) $1/\tau = 1$ and (b) $1/\tau = 25$ 
meV.
The curves correspond to the frequencies $\nu = 0.2 $ meV (solid),
$\nu = 1.0 $ meV (dotted), $\nu = 2.5 $ meV (dashed),
$\nu = 5.0 $ meV (dot-dashed), and $\nu = 10 $ meV (short-dashed-long-dashed).
The effect of finite frequency is more significant in the clean limit.

\item Fig. 4.
$\s2$ vs the impurity scattering rate $1/\tau$, for various frequencies.
The results are in the BCS limit, with $T/T_c = 0.1$. The curves correspond to
$\nu = 0$ (solid), 1 (dotted), 5 (dashed), and 10 meV (dot-dashed).

\item Fig. 5.
$\s2$ vs. frequency as calculated using an electron-phonon spectrum
with $\lambda = 1$ and $T_c = 29$ K as described in the text.
In (a) we use $1/\tau = 0$ (clean limit) and plot the results for various
temperatures, as indicated. The solid (dotted) curve is for the superconducting
(normal) state at $T/T_c = 0.99$. Clearly there is very little
difference throughout the frequency scale shown. Note the lack of a
cusp-like minimum at $\nu = 2 \Delta_0 \approx 10.4 $ meV. In (b) we
focus on the low temperature results ($T/T_c = 0.2$) and show the effect
of a small amount of impurity scattering. The cusp-like minimum at
$2 \Delta$ is clearly present.

\item Fig. 6.

$\s2$ vs. frequency for various coupling strengths, as indicated.
The results are for low temperature $T/T_c = 0.3$ and (a) relatively
clean, $1/\tau = 2 $ meV, and (b) $1/\tau = 25$ meV. 
The Coulomb repulsion $\mu^\ast$ has been adjusted to keep $T_c$ fixed
while the coupling strength has changed (simply by scaling the 
$\alpha^2F(\omega)$ spectrum).
Increased coupling strength suppresses $\s2$ and broadens the minimum
at $2\Delta$. Note that $2\Delta$ increases slightly as the coupling 
strength is increased.

\item Fig. 7.
$\s2$ vs. frequency for $1/\tau = 50$ meV.
The other parameters are the same
as used in Fig. 5. 
The various curves are for different temperatures,
$T/T_c = 0.99$ (solid --- superconducting, dotted --- normal),
$T/T_c = 0.5$ (dashed), and $T/T_c = 0.2$ (long-dashed).
Both the first pair and the second pair of curves differ from one another
only at very low frequency. Also shown is the normal state result at 
$T/T_c = 10$.

\item Fig. 8.
$\s2$ vs. frequency for $1/\tau = 0$ (clean limit) in the normal state, 
using the same parameters for the spectrum as in Fig. 5.
The solid curves are results of complete Eliashberg
calculations whereas the dotted curves are for the derived low
frequency Drude fits to the complete results.  Curves are labelled
by their reduced temperatures.

\item Fig. 9.
The measured $\s2$ vs. frequency at $T = 9$ K and at
$T = 300$ K (solid curves). Also shown are the theoretical fits, 
using the $\alpha^2F(\omega)$ described in the text, but with $\lambda = 0.2$
(dashed curves).
$T_c$ is kept fixed to the experimental value with a negative $\mu^\ast$.
Finally, theoretical fits are also shown with $\lambda = 1$ (dotted curves).
The latter curves are clearly incompatible with the experimental results.
\end{itemize}


\begin{thebibliography} {999}
\bibitem{26}
In the Drude model, $\sigma_2(\nu) = {(\nu \tau)^2 \over 1 + (\nu \tau)^2 }$.
This is what we mean when we use the term ``inverted Drude-like.''
\bibitem{30}
We have shown that the renormalization is actually $1/(1 + \lt)$. However,
$lim_{T \rightarrow 0} \lt (T) = \lambda$.
\end{thebibliography}
\end{document}